\documentclass[prb,aps,twocolumn,showpacs,floats]{revtex4}
\usepackage{psfig}

\begin{document}

\title{Quantum periodic cluster methods for strongly correlated electron systems}

\author{Tran Minh-Tien}
\affiliation{Asia Pacific Center for Theoretical Physics, POSTECH, Pohang, Republic of Korea, \\
Institute of Physics and Electronics, Vietnamese Academy of Science and Technology,
Hanoi, Vietnam.
}

%\maketitle
%\date{\today}

\pacs{71.10.-w, 71.10.Fd, 71.27.+a, 71.10.Pm}

\begin{abstract}
Quantum periodic cluster methods for strongly correlated electron systems
are reformulated and developed. The reformulation and development are based on 
a canonical transformation which periodizes the fermions in the cluster space. 
The dynamical cluster approximation
and the cellular dynamical mean field theory are related
each other
through the canonical transformation.
A cluster perturbation theory with periodic boundary
conditions is  developed.
It is found that the periodic cluster perturbation
theory converges rapidly with corrections $\mathcal{O}(1/L_c^2)$, where
$L_c$ is the linear size of the clusters, whereas the ordinary cluster
perturbation theory converges with corrections $\mathcal{O}(1/L_c)$. 
\end{abstract}

\maketitle

\section{Introduction}
One of the most active areas in the condensed matter physics is the search for
new methods which could capture essential features of electron correlations
and disorder in a controllable manner. Perhaps the most successful and widely
used methods are the dynamical mean field theory (DMFT)\cite{Metzner,GKKR} 
and the coherent potential approximation (CPA).\cite{cpa1,cpa2,cpa3} 
Both these methods are exact in infinite dimensions. 
However, at finite dimensions they neglect nonlocal correlations. This limitation has raised interests
in developing of quantum cluster methods which could capture nonlocal correlations. 
The most successful of self consistent quantum cluster methods 
are the molecular coherent potential approximation (MCPA) for disorder systems,\cite{mcpa}
and the cellular dynamical mean field theory (CDMFT)\cite{Kotliar} and the dynamical
cluster approximation (DCA)\cite{Jarrell, Jarrell1} for 
correlated electron systems. In these quantum cluster theories
the system lattice is split into a series of clusters.
Both local and nonlocal
correlations within the clusters are treated exactly, whereas the nonlocal
correlations between different clusters are treated in a mean-field approximation.
The mean field is taken into account by mapping the lattice problem onto a self-consistent
effective cluster problem. 
The MCPA and CDMFT are formulated on the real space,
and share a common microscopic definition. 
Within the CDMFT (or the MCPA) the cluster Green function  
is calculated with open boundary conditions. 
The DCA is traditionally formulated in the reciprocal space.
It is based on the idea of discretizing irreducible quantities on the reciprocal space.
Within the DCA the cluster Green function is calculated with periodic boundary conditions.
Due to the difference of boundary conditions it seems that the DCA and the CDMFT
(or the MCPA) have different microscopic definitions. For instance, the CDMFT
can be formulated within the self-energy functional approach, whereas 
it seems that the DCA
cannot be.\cite{Potthoff,Potthoff1} 
There is also a view of the DCA in the real space representation which shows 
a relation between the DCA and CDMFT.\cite{Biroli,Aryan,Biroli1} The view is based
on an unitary transformation of the hopping matrix. However, the interaction part 
still remains unchanged.  
In this paper we show that the DCA and the
CDMFT share a common microscopic definition through a canonical transformation.
The canonical transformation is similar to the unitary transformation which
was used to view the DCA in the real space. 
It periodizes the fermions in the cluster space, that the cluster
Green function has periodic boundary conditions. 
We apply the canonical
transformation to the whole Hamiltonian, and in the presence of a constrain
which prevents the umklapp momentum transfer from the superlattice to the cluster
space 
we obtain a periodized
Hamiltonian. The umklapp momentum transfer appears when the sum of two momenta of the
superlattice is beyond the first Brillouin zone, and is solely due to the 
imaginary superlattice construction of clusters with periodic boundary conditions.
Within the periodized Hamiltonian the DCA can be derived from the CDMFT.
In the such way, the DCA and the CDMFT can be unified into a common microscopic background.
The derivation also gives the
microscopic background of the discretization of irreducible quantities on
the reciprocal space, and clarifies the approximation nature of the DCA. 

The present paper consists of two parts. In the first part a derivation of the
DCA from the CDMFT through the canonical transformation for fermions is
presented. 
The second part is concerned with  the cluster perturbation
theory (CPT).\cite{Gros,Senechal,Senechal1}
The CPT is also a quantum cluster approach. However, in difference with the CDMFT
or the DCA, the CPT does not have the self consistency. The CPT can be viewed 
as the first term of a systematic expansion around strong coupling. However,
within the CPT the cluster Green function is calculated with open boundary
conditions, that the wave vector within a cluster is no longer a conserved
quantum number. There are also several approaches which generate periodic boundary conditions
for the CPT. One is to add appropriate hopping terms and then subtract them within strong
coupling perturbation theory.\cite{Dahnken} However, this approach give results which are
less accurate than the ones from open boundary conditions.\cite{Senechal1}
The other approach adapts the periodicity of the clusters from the DCA. It was used
to study the Ising model.\cite{Jarrell1}
In this paper we present the periodic cluster perturbation
theory for fermion systems. The periodic cluster perturbation
theory (PCPT) is the CPT formulated for the periodized fermions which are 
obtained after making the canonical transformation. The PCPT has truly the small 
parameter $1/L_c$, where $L_c$ is the cluster linear size. It turns out that
the PCPT converges quadratically in $1/L_c$, whereas the CPT converges linearly
in $1/L_c$. This is also tested on the study of an exact solvable model. 

The plan of the present paper is as follows. In Sec.~II we present the canonical
transformation which transforms the original fermions onto periodized ones.
In Sec.~III  we derive the DCA from the CDMFT after making the canonical transformation.
The PCPT is presented in Sec.~IV. In this section we also study an exact solvable
model in both the direct and periodized forms. Finally, the conclusion and remarks 
are presented in Sec.~V.

\section{Periodized fermions}
We present a periodization for fermions through
a canonical transformation of fermionic operators. For an illustration 
a fairly general model of correlated electron systems is
considered.  
The Hamiltonian of the model reads
\begin{eqnarray}
H &=& \sum_{i,j,\sigma} t(i,j) c^{\dagger}_{i\sigma}c^{\null}_{j\sigma} 
\nonumber \\
&&  + \sum_{i,j}\sum_{\{\sigma\}} 
U^{\{\sigma\}}(i,j) c^{\dagger}_{i\sigma_1}
c^{\null}_{i\sigma_2}c^{\dagger}_{j\sigma_3}c^{\null}_{j\sigma_4}, 
\label{h}
\end{eqnarray}
where $\{\sigma\}=\{\sigma_1,\sigma_2,\sigma_3,\sigma_4\}$. 
$c^{\dagger}_{i\sigma}$ ($c^{\null}_{i\sigma}$) is the creation (annihilation)
operator for fermion with spin $\sigma$ at lattice site $i$. 
$t(i,j)$ is the hopping integral, and $U^{\{\sigma\}}(i,j)$
is the interaction strength. We will consider a hypercubic lattice of linear size $L$
with the lattice constant $a=1$
on $d$ dimensions. The number of lattice sites thus is $N=L^d$.
The lattice is divided into a set of identical clusters of linear size $L_c$. The number of lattice
sites inside a cluster thus is $N_c=L_c^d$. The set of the clusters form a superlattice.
We use letters $i$, $j$,... to label the
lattice site of the original lattice, and use $\mathbf{r}_{i}$
to denote the position of site $i$. Letters 
$\mathbf{k}$, $\mathbf{p}$, ... are used to
denote the wave vectors of the reciprocal space of the original lattice.
Capital letters $I$, $J$,... are used to label the origin of the clusters. 
The origin coordinate of clusters are denoted by $\mathbf{R}_{I}$, and
the wave vectors of the reciprocal
space of the superlattice are denoted by
$\mathbf{K}$, $\mathbf{P}$, ... The lattice sites
inside a cluster are labeled by $a$, $b$,... Their position is
denoted by $\mathbf{\bar{r}}_a$, $\mathbf{\bar{r}}_b$,... and the wave
vectors of the reciprocal
space of the cluster are denoted by $\mathbf{\bar{k}}$, $\mathbf{\bar{p}}$,... 
Thus $\mathbf{r}_i=\mathbf{R}_I+\mathbf{\bar{r}}_a$, 
$\mathbf{k}=\mathbf{K}+\mathbf{\bar{k}}$,...
Note that $\exp(i \mathbf{\bar{k}} \cdot \mathbf{R})=1$,
since $\bar{k}_\alpha=2 \pi n_\alpha/L_c$,
and $R_\alpha=m_\alpha L_c$, where $n_\alpha$, $m_\alpha$ are integer.
We use the Greek letter $\sigma$ to denote the spin
(or other quantum) variables.  
With these notations
we rewrite the operators and model parameters as $c_{i\sigma} \equiv c_{a\sigma}(I)$, 
$t(i,j)\equiv t_{ab}(I,J)$, 
$U^{\{\sigma\}}(i,j)\equiv 
U_{ab}^{\{\sigma\}}(I,J)$,... 
Denoting $c^{\dagger}_{\mathbf{k}\sigma}$, $c^{\null}_{\mathbf{k}\sigma}$ 
the Fourier transform of $c^{\dagger}_{i\sigma}$, 
$c^{\null}_{i\sigma}$,  
the Hamiltonian (\ref{h}) can be rewritten in the reciprocal space
representation
\begin{eqnarray}
H = \sum_{\mathbf{k},\sigma} t(\mathbf{k}) c^{\dagger}_{\mathbf{k}\sigma}
c^{\null}_{\mathbf{k}\sigma} \hspace{5cm} 
\nonumber \\
 +\frac{1}{N} \sum_{\mathbf{k},\mathbf{k'},\mathbf{p}}\sum_{\{\sigma\}} 
U^{\{\sigma\}}(\mathbf{p}) c^{\dagger}_{\mathbf{k}+\mathbf{p},\sigma_1}
c^{\null}_{\mathbf{k},\sigma_2}c^{\dagger}_{\mathbf{k'-p},\sigma_3}
c^{\null}_{\mathbf{k'},\sigma_4}, 
\label{hp}
\end{eqnarray}
where $t(\mathbf{k})$ and $U^{\{\sigma\}}(\mathbf{p})$ are the Fourier transformation
of $t(i,j)$ and $U^{\{\sigma\}}(i,j)$, respectively, i.e.,
\begin{eqnarray}
t(i,j) &=& \frac{1}{N} \sum_{\mathbf{k}}
t(\mathbf{k})  
e^{i \mathbf{k}\cdot(\mathbf{r}_i-\mathbf{r}_j)}  , \\
U^{\{\sigma\}}(i,j) &=&
\frac{1}{N} \sum_{\mathbf{k}}
U^{\{\sigma\}}(\mathbf{k}) 
e^{i \mathbf{k}\cdot(\mathbf{r}_i-\mathbf{r}_j)} . 
\end{eqnarray} 

We introduce operators
\begin{eqnarray}
\bar{c}_{a\sigma}(\mathbf{K}) &=& \sqrt{\frac{N_c}{N}} 
\sum_{I} c_{a\sigma}(I) e^{- i \mathbf{K}\cdot(\mathbf{R}_I+\mathbf{\bar{r}}_a)} , 
\label{t1a} \\
\bar{c}_{a\sigma}^{\dagger}(\mathbf{K}) &=& \sqrt{\frac{N_c}{N}} 
\sum_{I} c_{a\sigma}^{\dagger}(I) e^{i \mathbf{K}\cdot(\mathbf{R}_I+\mathbf{\bar{r}}_a)} .
\label{t1b}
\end{eqnarray}
One can check that these operators satisfy the anticommutation relations
\begin{eqnarray*}
\{\bar{c}_{a\sigma}(\mathbf{K}),\bar{c}_{b\sigma'}^{\dagger}(\mathbf{P})\}
&=& \delta_{ab}\delta_{\sigma\sigma'}\delta(\mathbf{K}-\mathbf{P}) , \\
\{\bar{c}_{a\sigma}(\mathbf{K}),\bar{c}_{b\sigma'}(\mathbf{P})\} &=& 0 , \\
\{\bar{c}_{a\sigma}^{\dagger}(\mathbf{K}),\bar{c}_{b\sigma'}^{\dagger}(\mathbf{P})\} &=& 0 ,
\end{eqnarray*}
where $\{A,B\}=A B + B A$.
Since the anticommutation relations are preserved, the transformation 
(\ref{t1a})-(\ref{t1b}) is canonical.  
The inverse transformation reads
\begin{eqnarray}
c_{a\sigma}(I) &=& \sqrt{\frac{N_c}{N}} 
\sum_{\mathbf{K}} \bar{c}_{a\sigma}(\mathbf{K}) e^{i \mathbf{K}\cdot(\mathbf{R}_I+\mathbf{\bar{r}}_a)} , 
\label{t2a} \\
c_{a\sigma}^{\dagger}(I) &=& \sqrt{\frac{N_c}{N}} 
\sum_{\mathbf{K}} \bar{c}_{a\sigma}^{\dagger}(\mathbf{K}) e^{-i \mathbf{K}\cdot 
(\mathbf{R}_I+\mathbf{\bar{r}}_a)}  .
\label{t2b}
\end{eqnarray}
We also introduce other operators in the real superlattice space
\begin{eqnarray}
\bar{c}_{a\sigma}(I) &=& \sqrt{\frac{N_c}{N}} 
\sum_{\mathbf{K}} \bar{c}_{a\sigma}(\mathbf{K}) e^{i \mathbf{K}\cdot\mathbf{R}_I} , 
\label{t3a} \\
\bar{c}_{a\sigma}^{\dagger}(I) &=& \sqrt{\frac{N_c}{N}} 
\sum_{\mathbf{K}} \bar{c}_{a\sigma}^{\dagger}(\mathbf{K}) e^{-i \mathbf{K}\cdot\mathbf{R}_I} .
\label{t3b}
\end{eqnarray}
One can check that these operators (\ref{t3a})-(\ref{t3b}) also satisfy the 
fermionic anticommutation relations.
They annihilate or create a fermion at site $(I,a)$.
The transformation (\ref{t3a})-(\ref{t3b}) can be viewed as the Fourier transformation
for the fermion operators $\bar{c}_{a\sigma}(\mathbf{K})$, 
$\bar{c}_{a\sigma}^{\dagger}(\mathbf{K})$ in the superlattice space.
From Eqs. (\ref{t3a})-(\ref{t3b}) and (\ref{t1a})-(\ref{t1b}), one can show that
\begin{eqnarray}
\bar{c}_{a\sigma}(I) &=& \sqrt{\frac{1}{N}} 
\sum_{\mathbf{K},\mathbf{\bar{k}}} 
c_{\mathbf{K}+\mathbf{\bar{k}},\sigma} 
e^{i \mathbf{K}\cdot\mathbf{R}_I + i \mathbf{\bar{k}}\cdot\mathbf{\bar{r}}_a  } , 
\label{t4a} \\
\bar{c}_{a\sigma}^{\dagger}(I) &=& \sqrt{\frac{1}{N}} 
\sum_{\mathbf{K},\mathbf{\bar{k}}} 
c_{\mathbf{K}+\mathbf{\bar{k}},\sigma}^{\dagger} 
e^{- i \mathbf{K}\cdot\mathbf{R}_I - i \mathbf{\bar{k}}\cdot\mathbf{\bar{r}}_a  }.
\label{t4b}
\end{eqnarray}
In deriving Eqs.~(\ref{t4a})-(\ref{t4b}) we have used 
$\exp(i \mathbf{\bar{k}}\cdot\mathbf{R}_I)=1$. 
Since $\exp(i \mathbf{\bar{k}}\cdot\mathbf{\bar{r}}_{a+L_c})=
\exp(i \mathbf{\bar{k}}\cdot\mathbf{\bar{r}}_{a})$, one can see that
the fermionic operators $\bar{c}_{a\sigma}(I)$, $\bar{c}_{a\sigma}^{\dagger}(I)$ 
are periodic in the indexes $a$, whereas the original ones 
$c^{\null}_{a\sigma}(I)$, $c_{a\sigma}^{\dagger}(I)$ are not.
The fermion created by $\bar{c}_{a\sigma}^{\dagger}(I)$ is periodic in 
both the superlattice and cluster spaces. 
Due to the periodic property we call $\bar{c}_{a\sigma}(I)$, $\bar{c}_{a\sigma}^{\dagger}(I)$
the periodized fermion operators. 
The periodicity allows to perform
the Fourier transformation of $\bar{c}_{a\sigma}(I)$, $\bar{c}_{a\sigma}^{\dagger}(I)$
in the superlattice and cluster spaces
\begin{eqnarray}
\bar{c}_{\mathbf{\bar{k}}\sigma}(\mathbf{K})&=& \sqrt{\frac{1}{N}} 
\sum_{I,a} 
\bar{c}_{a\sigma}(I) 
e^{- i \mathbf{K}\cdot\mathbf{R}_I - i \mathbf{\bar{k}}\cdot\mathbf{\bar{r}}_a  } , 
\label{t4ia} \\
\bar{c}_{\mathbf{\bar{k}}\sigma}^{\dagger}(\mathbf{K}) &=& \sqrt{\frac{1}{N}} 
\sum_{I,a} 
\bar{c}_{a\sigma}^{\dagger}(I) 
e^{i \mathbf{K}\cdot\mathbf{R}_I + i \mathbf{\bar{k}}\cdot\mathbf{\bar{r}}_a  }.
\label{t4ib}
\end{eqnarray}
We want to emphasise here that both $\mathbf{K}$ and $\mathbf{\bar{k}}$  should
be restricted to the first Brillouin zones of the superlattice and cluster spaces.
From Eqs.~(\ref{t4a})-(\ref{t4b}) we obtain immediately 
$\bar{c}^{\null}_{\mathbf{\bar{k}}\sigma}(\mathbf{K})=
c^{\null}_{\mathbf{\bar{k}}+\mathbf{K},\sigma}$,
$\bar{c}^{\dagger}_{\mathbf{\bar{k}}\sigma}(\mathbf{K})=
c^{\dagger}_{\mathbf{\bar{k}}+\mathbf{K},\sigma}$.
However, these relations are valid only for momenta in the first Brillouin zones.
Thus, from the Hamiltonian
(\ref{hp}) we obtain
\begin{eqnarray}
H = \sum_{\mathbf{K},\mathbf{\bar{k}},\sigma} t(\mathbf{K}+\mathbf{\bar{k}}) 
\bar{c}^{\dagger}_{\mathbf{\bar{k}}\sigma}(\mathbf{K})
\bar{c}^{\null}_{\mathbf{\bar{k}}\sigma}(\mathbf{K}) \hspace{3cm}
\nonumber \\
 +\frac{1}{N} \sum_{\mathbf{K},\mathbf{K'},\mathbf{P}}
\sum_{\mathbf{\bar{k}},\mathbf{\bar{k}'},\mathbf{\bar{p}}}\sum_{\{\sigma\}} 
U^{\{\sigma\}}(\mathbf{P}+\mathbf{\bar{p}}) 
\bar{c}^{\dagger}_{\mathbf{\bar{k}}+\mathbf{\bar{p}},\sigma_1}(\mathbf{K}+\mathbf{P})
\nonumber \\
\bar{c}^{\null}_{\mathbf{\bar{k}},\sigma_2}(\mathbf{K})
\bar{c}^{\dagger}_{\mathbf{\bar{k}'}-\mathbf{\bar{p}},\sigma_3}(\mathbf{K'}-\mathbf{P})
\bar{c}^{\null}_{\mathbf{\bar{k}'},\sigma_4}(\mathbf{K'}). \hspace{.4cm}
\label{hpn}
\end{eqnarray}
Note that in Hamiltonian (\ref{hpn}) all sums over momenta are restricted to the
first Brillouin zones. Moreover, when $\mathbf{K}+\mathbf{P}$ and $\mathbf{K'}-\mathbf{P}$
are beyond the first Brillouin zone, one has to translate them back to the
first Brillouin zone, thus the momentum sums still are over the full first Brillouin
zone. Equivalently, 
the momenta of the periodized fermion operators take values 
$K_\alpha=2 \pi \text{mod}(n_\alpha,L/L_c)/L-\pi/L_c$.
In deriving the Hamiltonian (\ref{hpn}) we have used the following restriction
\begin{eqnarray}
\delta(\mathbf{p}+\mathbf{k}-\mathbf{q}) &\equiv&
\delta(\mathbf{P}+\mathbf{K}-\mathbf{Q}+\mathbf{\bar{G}})
\delta(\mathbf{\bar{p}}+\mathbf{\bar{k}}-\mathbf{\bar{q}}+\mathbf{\bar{G}}') 
\nonumber \\
&=&
\delta(\mathbf{P}+\mathbf{K}-\mathbf{Q}+\mathbf{\bar{G}})
\delta(\mathbf{\bar{p}}+\mathbf{\bar{k}}-\mathbf{\bar{q}}) ,
\label{const}
\end{eqnarray}
where $\mathbf{\bar{G}}$, $\mathbf{\bar{G}}'$ are arbitrary
momenta which are either zero or a vector of the reciprocal
superlattice space. $\mathbf{\bar{G}}$, $\mathbf{\bar{G}}'$ are also 
the momenta of the reciprocal cluster space. $\mathbf{\bar{G}} \neq 0$ just
means an umklapp process on the superlattice, and $\mathbf{\bar{G}}'$ can be
interpreted as a momentum transfered by the umklapp process from the superlattice
to the cluster space.  
The first line of Eq.~(\ref{const}) is exact, while the second line is a particular
restriction.
This restriction prevents the momentum transfer in the umklapp processes of the superlattice.
The umklapp processes appear when the
sum of two momenta of the superlattice is beyond the first Brillouin zone,
and they transfer a momentum modulo a vector of the reciprocal superlattice 
to the reciprocal cluster space that generate a momentum nonconservation.
However, the such momentum transfer is unphysical, it is solely due to the
imaginary discretization of the original lattice into superlattice of clusters
with periodic boundary conditions. 
In order to maintain the correct physics, one has to impose a constrain
which forbids the umklapp momentum transfer on the superlattice. 
The restriction (\ref{const}) is the constrain of the superlattice construction.
As we will see later on an example of exact solvable model, the constrain
indeed leads the periodized Hamiltonian
(\ref{hpn}) to the exact solution.
Performing 
the Fourier transformation (\ref{t4ia})-(\ref{t4ib}), from
the Hamiltonian (\ref{hpn}) we obtain
\begin{eqnarray}
H = \sum_{I,J,a,b,\sigma} \bar{t}_{ab}(I,J) \bar{c}^{\dagger}_{a\sigma}(I)
\bar{c}_{b\sigma}(J) + \hspace{2cm} 
\nonumber \\
 \sum_{I,J,a,b,\{\sigma\}} 
\bar{U}^{\{\sigma\}}_{ab}(I,J) 
\bar{c}^{\dagger}_{a\sigma_1}(I)
\bar{c}_{a\sigma_2}(I)\bar{c}^{\dagger}_{b\sigma_3}(J)
\bar{c}_{b\sigma_4}(J),
\label{h1} 
\end{eqnarray} 
where the hopping integral and interaction of the periodized fermions now are
\begin{eqnarray}
\bar{t}_{ab}(I,J) = \frac{1}{N} \sum_{\mathbf{K},\mathbf{\bar{k}}}
t(\mathbf{K}+\mathbf{\bar{k}}) 
e^{i \mathbf{K}\cdot(\mathbf{R}_I-\mathbf{R}_J) + i \mathbf{\bar{k}}\cdot
(\mathbf{\bar{r}}_a-\mathbf{\bar{r}}_b) } ,\hspace{0.2cm} 
\label{hop}\\
\bar{U}^{\{\sigma\}}_{ab}(I,J) = 
\frac{1}{N} \sum_{\mathbf{K},\mathbf{\bar{k}}}
U^{\{\sigma\}}(\mathbf{K}+\mathbf{\bar{k}}) e^{i \mathbf{K}\cdot(\mathbf{R}_I-\mathbf{R}_J) + 
 i \mathbf{\bar{k}}\cdot(\mathbf{\bar{r}}_a-\mathbf{\bar{r}}_b)}  . \nonumber \\
\label{int}
\end{eqnarray} 
One can notice that
$\bar{t}_{ab}(I,J)$, $\bar{U}^{\{\sigma\}}_{ab}(I,J)$ are cyclic in the indexes
$a$, $b$, whereas $t_{ab}(I,J)$, $U^{\{\sigma\}}_{ab}(I,J)$ are not. This means
that the Fourier transformation of $\bar{t}_{ab}(I,J)$ and 
$\bar{U}^{\{\sigma\}}_{ab}(I,J)$ are diagonal in the reciprocal space.
Denoting
$\bar{t}_{\mathbf{\bar{k}}}(\mathbf{K})$, 
$\bar{U}^{\{\sigma\}}_{\mathbf{\bar{k}}}(\mathbf{K})$  the Fourier
transformation of
$\bar{t}_{ab}(I,J)$, $\bar{U}^{\{\sigma\}}_{ab}(I,J)$, one can see
immediately from Eqs.~(\ref{hop})-(\ref{int})  that $\bar{t}_{\mathbf{\bar{k}}}(\mathbf{K})=t(\mathbf{K}+\mathbf{\bar{k}})$,
$\bar{U}_{\mathbf{\bar{k}}}^{\{\sigma\}}(\mathbf{K})=
U^{\{\sigma\}}(\mathbf{K}+\mathbf{\bar{k}})$.
Note that the hopping term (\ref{hop}) was also obtained previously by employing
an unitary transformation.\cite{Biroli,Biroli1} The unitary transformation is similar to
the canonical transformation for fermion operators.  
Beside the periodicity, the periodized hopping
(\ref{hop}) and interaction (\ref{int}) are quite different from 
the original hopping and interaction, respectively. They may connect those lattice
sites that the original ones do not.
For instance, if the original hopping is nearest neighbor  then
the hopping only couples  the nearest neighbor sites either within the clusters,
or on the cluster boundaries of the nearest neighbor clusters.
However, in the same case the periodized hopping can couple 
a lattice site with any other lattice site. 
The Hamiltonian (\ref{h1}) is adequate to quantum cluster
methods with periodic boundary conditions.

Next, we consider the one-particle Green function of the 
periodized fermions
\begin{eqnarray}
\bar{G}_{ab}(\mathbf{K},z)= 
\ll \bar{c}_{a\sigma}(\mathbf{K}) | 
\bar{c}_{b\sigma}^{\dagger}(\mathbf{K}) \gg_{z} .
\end{eqnarray}
Without difficulty one can show that
\begin{eqnarray}
\bar{G}_{ab}(\mathbf{K}) = 
\frac{1}{N_c} \sum_{\mathbf{\bar{k}}} G(\mathbf{K}+\mathbf{\bar{k}})
e^{i \mathbf{\bar{k}}\cdot(\mathbf{\bar{r}}_a-\mathbf{\bar{r}}_b)}, 
\label{g1} 
\end{eqnarray}
where $G(\mathbf{k})=\ll c_{\mathbf{k}\sigma} | 
c_{\mathbf{k}\sigma}^{\dagger} \gg_{z} $ is the Green function of the original
fermions. Equation~(\ref{g1}) shows that the Green function of the periodized fermions
is diagonal in the reciprocal cluster space. 
We immediately obtain
\begin{eqnarray}
G(\mathbf{K}+\mathbf{\bar{k}},z) = \bar{G}_{\mathbf{\bar{k}}}(\mathbf{K},z) ,
\label{gg1} 
\end{eqnarray}
Equation~(\ref{gg1}) expresses the identity
of the Green functions of the original fermions and of the periodized fermions.
The Dyson equation of the Green function $\bar{G}_{ab}(\mathbf{K},z)$ reads
\begin{eqnarray}
	\big[\hat{G}(\mathbf{K},z)\big]^{-1} = z\hat{I} -\hat{\bar{t}}(\mathbf{K})
	- \hat{\Sigma}(\mathbf{K},z) ,
\end{eqnarray}
where the hat symbol denotes the matrix notation in  the cluster space.
$\hat{\Sigma}(\mathbf{K},z)$ is the self energy.
Since the Green function $\hat{G}(\mathbf{K},z)$ and the hopping 
$\hat{\bar{t}}(\mathbf{K})$ are diagonal in the reciprocal cluster space,
the self energy $\hat{\Sigma}(\mathbf{K},z)$ must diagonal in the
reciprocal cluster space too. This is consistent with the Hamiltonian
(\ref{h1}), where both the hopping and interaction terms are diagonal.

\section{Dynamical cluster approximation}
With the periodized Hamiltonian (\ref{h1}) we can derive the DCA from the
CDMFT (or from the MCPA).
Applying the CDMFT to the periodized fermions we obtain
\begin{eqnarray}
\hat{\bar{G}}(\mathbf{K},z)=\big[z \hat{I} - \hat{\bar{t}}(\mathbf{K}) - 
\hat{\bar{\Sigma}}(z) \big]^{-1} .
\label{s1} 
\end{eqnarray}
The hopping matrix $\hat{\bar{t}}(\mathbf{K})$ is the Fourier transformation
of the periodized hopping (\ref{hop}) in the superlattice space, i.e.,
\begin{eqnarray}
	\bar{t}_{ab}(\mathbf{K}) &=& \frac{1}{N_c} \sum_{\mathbf{\bar{k}}}
t(\mathbf{K}+\mathbf{\bar{k}}) 
e^{i \mathbf{\bar{k}}\cdot
(\mathbf{\bar{r}}_a-\mathbf{\bar{r}}_b) } . 
\label{hopdca}
\end{eqnarray}
The self energy $\hat{\bar{\Sigma}}(z)$ is determined from an effective single cluster
problem.
The action of the effective single cluster is
\begin{eqnarray}
S_{\text{eff}} &=& - \int_{0}^{\beta} \int_{0}^{\beta}
d\tau d\tau' \sum_{a,b,\sigma}
\bar{c}^{\dagger}_{a\sigma}(\tau) \bar{\cal{G}}^{-1}_{ab}(\tau-\tau') 
\bar{c}^{\null}_{b\sigma}(\tau')
 \nonumber \\
&& + \int_{0}^{\beta} d\tau \sum_{a,b,\{\sigma\}} \bar{U}_{ab}^{\{\sigma\}}  
(\bar{c}^{\dagger}_{a\sigma_1}
\bar{c}_{a\sigma_2}\bar{c}^{\dagger}_{b\sigma_3} 
\bar{c}_{b\sigma_4} ) (\tau) ,
\label{seff}
\end{eqnarray}
where $\bar{\cal{G}}^{-1}_{ab}(\tau-\tau')$ plays the role of the 
effective mean field acting on the
cluster. The interaction in the effective single cluster is
obtained from Eq.~(\ref{int})
\begin{eqnarray}
\bar{U}_{ab}^{\{\sigma\}} =
\frac{1}{N} \sum_{\mathbf{K},\mathbf{\bar{k}}}
U^{\{\sigma\}}(\mathbf{K}+\mathbf{\bar{k}})
e^{i \mathbf{\bar{k}}\cdot(\mathbf{\bar{r}}_a
- \mathbf{\bar{r}}_b) } .  
\label{ueff}
\end{eqnarray} 
The self consistency requires the identity of the cluster Green function
obtained from the effective single cluster problem (\ref{seff}) and the cluster
Green function of the original cluster, i.e.,
\begin{eqnarray}
\frac{N_c}{N} \sum_{\mathbf{K}}
\hat{\bar{G}}(\mathbf{K},z) = \big[ \hat{\bar{\cal{G}}}^{-1}(z) - \hat{\bar{\Sigma}}(z)
\big]^{-1} .
\label{s2}
\end{eqnarray}
Since in the reciprocal cluster space the lattice Green function $\hat{\bar{G}}(\mathbf{K},z)$ 
and the self energy
$\hat{\bar{\Sigma}}(z)$ are diagonal, the Green function
of the effective medium $\hat{\bar{\cal{G}}}(z)$ must be diagonal too.
After making the Fourier transformation, we obtain
from Eqs.~(\ref{s1}), (\ref{s2})  
\begin{eqnarray}
\bar{G}_{\mathbf{\bar{k}}}(\mathbf{K},z)=\frac{1}{z  - \bar{t}_{\mathbf{\bar{k}}}(\mathbf{K}) - 
\bar{\Sigma}_{\mathbf{\bar{k}}}(z)} , \\
\frac{N_c}{N} \sum_{\mathbf{K}}
\bar{G}_{\mathbf{\bar{k}}}(\mathbf{K},z) = 
\frac{1}{ \bar{\cal{G}}^{-1}_{\mathbf{\bar{k}}}(z) - \bar{\Sigma}_{\mathbf{\bar{k}}}(z)
} .
\label{ss}  
\end{eqnarray}
Since $\bar{t}_{\mathbf{\bar{k}}}(\mathbf{K})=t(\mathbf{\bar{k}}+\mathbf{K})$, 
$\bar{U}_{\mathbf{\bar{k}}}=(N_c/N) \sum_{\mathbf{K}}
U(\mathbf{K}+\mathbf{\bar{k}})$,
and $\bar{G}_{\mathbf{\bar{k}}}(\mathbf{K},z)=G(\mathbf{\bar{k}}+\mathbf{K},z)$
we recover the DCA equations. 
Equations~(\ref{s1})-(\ref{s2}) can be viewed as the DCA equations
formulated in the real cluster space. 
Originally, the DCA is formulated in the reciprocal space.\cite{Jarrell} It is based
on the idea of discretizing irreducible quantities on the reciprocal space.
Later, the DCA was also viewed in the real space by introducing an unitary transformation
for the hopping matrix.\cite{Biroli,Aryan,Biroli1} 
However, in this view only the hopping term is transformed,
whereas the interaction term remains unchanged. As a consequence, the hopping and 
interaction terms are not treated on an equal footing. It correctly recovers the DCA  
only for a local single-site interaction. In the presented approach both the hopping
and interaction terms are treated on an equal footing. It also clarifies 
why within the DCA the lattice quantities, for instance the hopping term 
in the lattice Green function (Eq.~(\ref{hopdca})), are unchanged, whereas 
the cluster quantities, for instance the cluster
interactions (Eq.~(\ref{ueff})), are discretized or coarse grained. 
Thus, the DCA can be used in
both the reciprocal and real spaces. In the real space the DCA is just the CDMFT
applied to the periodized Hamiltonian (\ref{h1}). 
Due to this formulation the DCA has  
the properties of the CDMFT. For instance, the DCA can also be formulated within 
the self-energy functional approach in the real space by the same way as of the CDMFT.\cite{Potthoff,Potthoff1} 
So far we have shown that both the DCA and CDMFT 
share a common microscopic definition. They are related each other through the
canonical periodization transformation.
An extensive discussion about the two methods
is already given in the recent review.\cite{Jarrell1}

\section{Periodic cluster perturbation theory}
The CPT is also a quantum cluster approach, although it does not have the
self consistency.\cite{Gros,Senechal,Senechal1} 
In this section we present a CPT with periodic boundary conditions.
We refer it as PCPT. Basically, the PCPT is the CPT 
applied to the periodized fermions. 
A similar PCPT which employs the formulation of the DCA for spin systems
was used previously to study the Ising model.\cite{Jarrell1}
We will consider the system with the nearest neighbor
hopping. In the periodization version  the hopping (\ref{hop}) is no longer only nearest neighbor.
It may couple a lattice site with any other lattice site.  
As the CPT, the PCPT splits the intra-
and inter-cluster parts of the hopping term. The intra-cluster
part of the Hamiltonian is
\begin{eqnarray}
H^{\text{c}} = \sum_{a,b,\sigma} \bar{t}_{ab} \bar{c}^{\dagger}_{a\sigma}
\bar{c}_{b\sigma} + 
 \sum_{a,b,\{\sigma\}} 
\bar{U}^{\{\sigma\}}_{ab} 
\bar{c}^{\dagger}_{a\sigma_1}
\bar{c}_{a\sigma_2}\bar{c}^{\dagger}_{b\sigma_3}
\bar{c}_{b\sigma_4},
\label{hc} 
\end{eqnarray}
where 
$\bar{t}_{ab} = \bar{t}_{ab}(I,I)$, and $\bar{U}^{\{\sigma\}}_{ab}=\bar{U}^{\{\sigma\}}_{ab}(I,I)$.
The inter-cluster part is
\begin{eqnarray}
H^{\text{ic}} = \sum_{I,J,a,b,\sigma} \delta\bar{t}_{ab}(I,J) \bar{c}^{\dagger}_{a\sigma}(I)
\bar{c}_{b\sigma}(J) , 
\label{hi} 
\end{eqnarray}
where $\delta\bar{t}_{ab}(I,J)=\bar{t}_{ab}(I,J)-\bar{t}_{ab}$. Here we have neglected
the inter-cluster interactions. Originally, the CPT is formulated for local single-site
interactions, thus the inter-cluster interactions are absent. Often quantum cluster approaches
either neglect the inter-cluster interactions or treat them in a static mean-field approximation.
However the inter-cluster interactions can be incorporated beyond the static mean-field approximation
by employing a self consistent approach.\cite{Smith,Maier1}
Note that the cluster hopping $\bar{t}_{ab}$ may couple any sites within the cluster.
At a finite cluster it is reduced by a factor of $\mathcal{O}(1/L_c^2)$.\cite{Jarrell1} 
The inter-cluster
hopping $\delta\bar{t} \sim 1/L_c$ for large linear cluster size $L_c$ compared to 
its counterpart $t$ of the original CPT.\cite{Jarrell1,Maier} 
The nature of the approximation of the CPT is a strong coupling expansion in which the small
parameter is the inter-cluster hopping.\cite{Gros,Senechal,Senechal1}
Therefore in the PCPT the small parameter is truly 
$1/L_c$. 

Within the PCPT the cluster Hamiltonian (\ref{hc}) is solved exactly. 
We obtain the Dyson equation for the
cluster Green function
\begin{eqnarray}
\bar{G}^{\text{c}}_{\mathbf{\bar{k}}}(z)=\frac{1}{z-\bar{t}_{\mathbf{\bar{k}}}-
\bar{\Sigma}^{\text{c}}_{\mathbf{\bar{k}}}(z)} ,
\label{cpt1}
\end{eqnarray} 
where $\bar{t}_{\mathbf{\bar{k}}}=(N_c/N)\sum_{\mathbf{K}} t(\mathbf{K}+\mathbf{\bar{k}})$, and $\bar{\Sigma}^{\text{c}}_{\mathbf{\bar{k}}}(z)$ is the cluster self energy. Here the Dyson 
equation is written in the reciprocal space representation. 
We can make the Fourier transformation in the cluster space because the fermions are
periodized in the cluster space, thus the cluster Green function and
self energy are diagonal in the reciprocal space. 
As the CPT, within the PCPT the self energy of the lattice Green function is approximated
by the cluster self energy. Thus we obtain 
\begin{eqnarray}
\bar{G}_{\mathbf{\bar{k}}}(\mathbf{K},z)=
\frac{1}{z-\bar{t}_{\mathbf{\bar{k}}}(\mathbf{K})-
\bar{\Sigma}^{\text{c}}_{\mathbf{\bar{k}}}(z)} .
\label{cpt2}
\end{eqnarray} 
Since $\bar{G}_{\mathbf{\bar{k}}}(\mathbf{K},z)=G(\mathbf{K}+\mathbf{\bar{k}},z)$
and $\bar{t}_{\mathbf{\bar{k}}}(\mathbf{K})=t(\mathbf{K}+\mathbf{\bar{k}})$ we
finally obtain the Green function of the original fermions
\begin{eqnarray}
G(\mathbf{K}+\mathbf{\bar{k}},z)=
\frac{1}{z-t(\mathbf{K}+\mathbf{\bar{k}})-
\bar{\Sigma}^{\text{c}}_{\mathbf{\bar{k}}}(z)} .
\label{cpt3}
\end{eqnarray}
Equations~(\ref{cpt1})-(\ref{cpt3}) are the principal equations of the PCPT. 
As the CPT,  the PCPT is exact in the limits 
$L_c \rightarrow \infty$, $U/t=0$, and  $t/U=0$ for local interactions. 
The PCPT is formulated
in the reciprocal space, hence it avoids the diagonalizations of the Green function
in both real and reciprocal spaces, that significantly reduces the computational
time compared to the CPT. Note that the present PCPT is different to the ordinary
periodic CPT,\cite{Senechal1,Dahnken} where the periodicity is taken into account
by including additional hoppings. In the ordinary periodic CPT the hopping is still
nearest neighbor, while in the present PCPT it is no longer only nearest neighbor.
Moreover, in the present PCPT all hoppings within a cluster are treated exactly, 
whereas in the ordinary periodic CPT the adding and subtracting hoppings are treated on 
different footings.

For an illustration we study the one dimensional large-$\mathcal{N}$ model,
originally
introduced by Affleck and Marston in two dimensions.\cite{Affleck}
The model was also used for comparison between 
the DCA and CDMFT schemes.\cite{Biroli,Aryan} 
We will compare the PCPT and CPT
schemes to its exact solution. 
The Hamiltonian of the model reads
\begin{eqnarray}
H &=& - t \sum_{<i,j>,\sigma} c^{\dagger}_{i\sigma} c^{\null}_{j\sigma} + 
\mu \sum_{i\sigma} c^{\dagger}_{i\sigma} c^{\null}_{i\sigma}
\nonumber \\
&& + \frac{J}{2 \mathcal{N}} \sum_{<i,j>,\sigma,\sigma'} 
c^{\dagger}_{i\sigma} c^{\null}_{i\sigma'} c^{\dagger}_{j\sigma'} c^{\null}_{j\sigma} ,
\label{ham}
\end{eqnarray}
where $<i,j>$ denotes the nearest neighbor sites, and $\sigma=1,...,\mathcal{N}$.
As usually, $t$ is the hopping integral, $J$ is the exchange strength, and $\mu$ is
the chemical potential.
We will consider the large $\mathcal{N}$ limit. In this limit the model can be solved
exactly. Indeed, the quantity 
$\chi=(1/\mathcal{N}) \sum_{\sigma} c^{\dagger}_{i,\sigma} c^{\null}_{i+1,\sigma} $
does not fluctuate, that the static mean field theory is exact. We obtain the
exact solution
\begin{eqnarray}
\chi^{\text{exact}} &=& \frac{1}{2L} \sum_{\mathbf{k}} f(\beta E(\mathbf{k})) \gamma(\mathbf{k}) , \\
E(\mathbf{k}) &=& -(t+J \chi^{\text{exact}}) \gamma(\mathbf{k}) + \mu ,
\end{eqnarray}
where $\beta$ is the inverse temperature, and $f(x)=1/(\exp(x)+1)$
is the Fermi function, and $\gamma(\mathbf{k})=2 \cos(k)$.    

\begin{figure}[t]
\begin{center}
\psfig{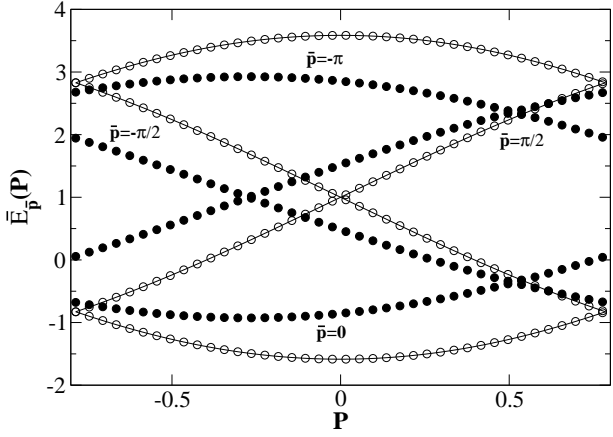}
\end{center}
\caption{\label{figa}
The dispersion $\bar{E}_{\mathbf{\bar{p}}}(\mathbf{P})$ of the periodized
fermions (the open symbols)
for the cluster size $L_c=4$ in comparison with the dispersion
$E(\mathbf{P}+\mathbf{\bar{p}})$ of the original fermions (the line).
The filled symbols are $\bar{E}^{\text{u}}_{\mathbf{\bar{p}}}(\mathbf{P})$
of the umklapp case for $L_c=4$.
 ($t=J=1$, $\mu=1$, $\beta=10$).}
\end{figure}
    
Within the CPT, the cluster of size $L_c$ has the following Hamiltonian
\begin{eqnarray}
H^{\text{c}} &=&  -t \sum_{a=1}^{L_c-1}\sum_{\sigma} c^{\dagger}_{a,\sigma} c^{\null}_{a+1,\sigma} +
\text{h.c.} +
\mu \sum_{a=1}^{L_c}\sum_{\sigma} c^{\dagger}_{a\sigma} c^{\null}_{a\sigma}
\nonumber \\
&& + \frac{J}{2 \mathcal{N}} \sum_{a=1}^{L_c-1}\sum_{\sigma,\sigma'} 
c^{\dagger}_{a,\sigma} c^{\null}_{a,\sigma'} c^{\dagger}_{a+1,\sigma'} c^{\null}_{a+1,\sigma}
+\text{h.c.} .
\label{hamcpt}
\end{eqnarray}
The cluster Hamiltonian (\ref{hamcpt}) can be solved exactly, for instance by the exact
diagonalization. However,
at the large-$\mathcal{N}$ limit it  can simply be
solved exactly. Indeed, the cluster quantity
$\chi_{a}^{\text{c}}=(1/\mathcal{N}) \sum_{\sigma} c^{\dagger}_{a,\sigma} c^{\null}_{a+1,\sigma}$
does not fluctuate, and the cluster Hamiltonian (\ref{hamcpt}) becomes quadratic and 
can simply be diagonalized.
However, in this case $\chi_{a}^{\text{c}}$ may depends on the cluster index.
The cluster Green function is
\begin{eqnarray}
\hat{G}^{\text{c}}(z) = \big[ (z - \mu) \hat{I} - \hat{t} - \hat{\Sigma}^{\text{c}} ]^{-1} ,
\label{gcpt}  
\end{eqnarray}
where $\hat{t}$ is the cluster hopping matrix, and $\hat{\Sigma}^{\text{c}}$ is the
cluster self energy
\begin{equation}
\Sigma^{\text{c}}_{ab} = -J (\chi_{a}^{\text{c}} \delta_{a,b-1}+\chi_{b}^{\text{c}} \delta_{a,b+1} ) .
\label{scpt}
\end{equation}
The cluster quantity $\chi_{a}^{\text{c}}$ can be calculated from the cluster Green
function
\begin{equation}
\chi_{a}^{\text{c}} = - \int \frac{d\omega}{\pi} f(\beta \omega) \text{Im}
G_{a+1,a}^{\text{c}}(\omega+i 0^{+}) .
\label{ccpt}  
\end{equation}
Equations (\ref{gcpt})-(\ref{ccpt}) give the exact solution of the cluster
Hamiltonian (\ref{hamcpt}) at the large-$\mathcal{N}$ limit. Once the exact
cluster solution is obtained, the lattice Green function within the CPT is
computed by\cite{Senechal,Senechal1}
\begin{equation}
G^{\text{CPT}}(\mathbf{k},z)=\frac{1}{L_c} \sum_{ab} G_{ab}(\mathbf{K},z) 
e^{-i \mathbf{k}\cdot(\mathbf{\bar{r}}_a-\mathbf{\bar{r}}_b)} ,
\label{lgcpt}
\end{equation}
where $G^{-1}_{ab}(\mathbf{K},z)=(z-\mu)\delta_{ab}-t_{ab}(\mathbf{K})-\Sigma^{\text{c}}_{ab}$. 
The lattice quantity 
$\chi=(1/\mathcal{N}) \sum_{\sigma} c^{\dagger}_{i,\sigma} c^{\null}_{i+1,\sigma} $
within the CPT is calculated from this lattice Green function 
\begin{eqnarray}
&& \chi^{\text{CPT}} =
-\int \frac{d\omega}{\pi} f(\beta \omega) \text{Im}
G^{\text{CPT}}_{i+1,i}(\omega+i 0^{+})
\nonumber \\
&=& - \frac{1}{L} \sum_{\mathbf{K},a}\int \frac{d\omega}{\pi} f(\beta \omega) \text{Im}
G_{a+1,a}(\mathbf{K},\omega+i 0^{+})  .
\label{cptchi}
\end{eqnarray}

\begin{figure}[t]
\begin{center}
\psfig{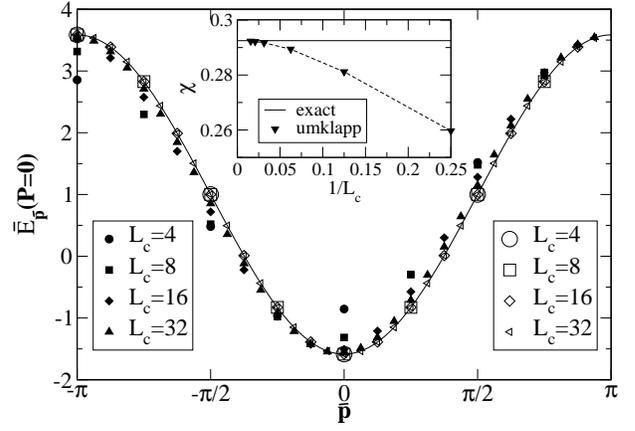}
\end{center}
\caption{\label{figb}
The dispersion $\bar{E}_{\mathbf{\bar{p}}}(\mathbf{P}=0)$ of the periodized
fermions for various cluster sizes $L_c$ (the open symbols) in comparison with
$E(\mathbf{P}+\mathbf{\bar{p}})$ of the original fermions (the line)
at $\mathbf{P}=0$. The filled symbols are 
$\bar{E}^{\text{u}}_{\mathbf{\bar{p}}}(\mathbf{P}=0)$ of the umklapp case. The
inset plots $\chi^{\text{u}}$ of the umklapp case in comparison with the
exact value.
($t=J=1$, $\mu=1$, $\beta=10$).}
\end{figure}

Next, we consider the one dimensional large-$\mathcal{N}$ model 
(\ref{ham}) in the periodization form.
The periodized Hamiltonian reads
\begin{eqnarray}
H = \sum_{I,J,a,b,\sigma} \bar{t}_{ab}(I,J)
\bar{c}^{\dagger}_{a\sigma}(I) \bar{c}^{\null}_{b\sigma}(J) +
\mu \sum_{I,a,\sigma} \bar{c}^{\dagger}_{a\sigma}(I) \bar{c}^{\null}_{a\sigma}(I)
 \nonumber \\
+\frac{1}{2 \mathcal{N}} \sum_{I,J,a,b,\sigma,\sigma'} \bar{J}_{ab}(I,J) 
\bar{c}^{\dagger}_{a\sigma}(I) \bar{c}^{\null}_{a\sigma'}(I) 
\bar{c}^{\dagger}_{b\sigma'}(J) \bar{c}^{\null}_{b\sigma}(J), \hspace{0.2cm}
\label{pham} 	
\end{eqnarray}
where $\bar{t}_{ab}(I,J)$ and $\bar{J}_{ab}(I,J)$ are determined
by the periodization transformation (\ref{hop}) and (\ref{int}), respectively.
The Hamiltonian (\ref{pham}) can be solved exactly in the large-$\mathcal{N}$ limit.
Introducing quantity 
$\bar{\chi}_{\mathbf{\bar{p}}}(\mathbf{P})=(1/\mathcal{N}) \sum_{\sigma}
\bar{c}^{\dagger}_{\mathbf{\bar{p}}\sigma}(\mathbf{P}) 
\bar{c}^{\null}_{\mathbf{\bar{p}}\sigma}(\mathbf{P}) $ we obtain 
\begin{eqnarray}
\bar{\chi}_{\mathbf{\bar{p}}}(\mathbf{P}) &=& 
f(\beta \bar{E}_{\mathbf{\bar{p}}}(\mathbf{P})) , 
\label{ec1} \\
\bar{E}_{\mathbf{\bar{p}}}(\mathbf{P}) &=& -t \gamma(\mathbf{P}+\mathbf{\bar{p}}) + \mu
\nonumber \\
&& -J \frac{1}{L} \sum_{\mathbf{K},\mathbf{\bar{k}}}
 \gamma(\mathbf{P}+\mathbf{\bar{p}}-\mathbf{K}-\mathbf{\bar{k}})\bar{\chi}_{\mathbf{\bar{k}}}(\mathbf{K}) . 	
\label{ec2}
\end{eqnarray}
When Eqs.~(\ref{ec1})-(\ref{ec2}) are solved, the quantity 
$\chi=(1/\mathcal{N}) \sum_{\sigma} c^{\dagger}_{i,\sigma} c^{\null}_{i+1,\sigma} $
of the original fermions can be calculated too
\begin{eqnarray}
	\chi^{\text{P}}= \frac{1}{2L} \sum_{\mathbf{P},\mathbf{\bar{p}}}
	f(\beta \bar{E}_{\mathbf{\bar{p}}}(\mathbf{P})) \gamma(\mathbf{P}+\mathbf{\bar{p}}).
\end{eqnarray}
Numerically, we alway obtain $\chi^{\text{P}}=\chi^{\text{exact}}$ independently on the
cluster size $L_c$. Moreover, since the Green function
$\ll \bar{c}_{\mathbf{\bar{k}}}(\mathbf{K}) | \bar{c}^{\dagger}_{\mathbf{\bar{k}}}(\mathbf{K}) \gg 
=\ll c(\mathbf{K}+\mathbf{\bar{k}}) | c^{\dagger}(\mathbf{K}+\mathbf{\bar{k}}) \gg $,
one must have the identity of the dispersions of periodized and original fermions $\bar{E}_{\mathbf{\bar{p}}}(\mathbf{P})=E(\mathbf{P}+\mathbf{\bar{p}})$. 
In Fig.~\ref{figa} we plot the dispersions for the cluster size $L_c=4$.
It confirms the identity of the dispersions of periodized and original fermions.
In Fig.~\ref{figb} we plot the dispersions at fixed $\mathbf{P}=0$ for various cluster
sizes $L_c$. It shows the identity of the dispersions of periodized and original fermions
is independent on the cluster size. 
These results confirm the equivalent of the periodized
and original Hamiltonians. 
Next, we consider the umklapp case where
the constrain (\ref{const}) is relaxed by its exact relation. Then we obtain
\begin{eqnarray}
\bar{\chi}^{\text{u}}_{\mathbf{\bar{p}}}(\mathbf{P}) &=& 
f(\beta \bar{E}^{\text{u}}_{\mathbf{\bar{p}}}(\mathbf{P})) , 
\label{uec1} \\
\bar{E}^{\text{u}}_{\mathbf{\bar{p}}}(\mathbf{P}) 
&=& -t \gamma(\mathbf{P}+\mathbf{\bar{p}}) + \mu
\nonumber \\
&&\hspace{-1cm} -J \frac{1}{L} \sum_{\mathbf{K},\mathbf{\bar{k}}}
 \gamma(\mathbf{P}+\mathbf{\bar{p}}-\mathbf{K}-\mathbf{\bar{k}}+\mathbf{\bar{G}})
 \bar{\chi}^{\text{u}}_{\mathbf{\bar{k}}}(\mathbf{K}),
\label{uec2}
\end{eqnarray}
where $\mathbf{\bar{G}}$ is the umklapp transfer momentum. 
Since $|\mathbf{\bar{G}}| \sim 1/L_c$, in the large $L_c$ limit this umklapp
case approaches to the normal one.
The lattice quantity $\chi=(1/\mathcal{N}) \sum_{\sigma} c^{\dagger}_{i,\sigma} 
c^{\null}_{i+1,\sigma} $
of the original fermions can be calculated by 
\begin{eqnarray}
	\chi^{\text{u}}= \frac{1}{2L} \sum_{\mathbf{P},\mathbf{\bar{p}}}
	f(\beta \bar{E}^{\text{u}}_{\mathbf{\bar{p}}}(\mathbf{P})) \gamma(\mathbf{P}+\mathbf{\bar{p}}).
\end{eqnarray}
In numerical calculations we take $\mathbf{\bar{G}}=2 \pi/L_c$.
It turns out numerically that $\chi^{\text{u}}$ is never equal to $\chi^{\text{exact}}$, although it
approaches to the exact value as $1/L_c^2$, as shown in the inset of Fig.~\ref{figb}.
In Fig.~\ref{figa} and \ref{figb} we plot also the dispersion 
$\bar{E}^{\text{u}}_{\mathbf{\bar{p}}}(\mathbf{P})$. These figures show that the dispersion
of the periodized fermions in the presence of the umklapp momentum transfer is not exact.
These results confirm that the constrain is important in order to maintain the correct
physics.

Now we apply the PCPT to the one dimensional large-$\mathcal{N}$ model
(\ref{ham}).
The cluster periodized Hamiltonian is obtained from (\ref{pham})
\begin{eqnarray}
H^{\text{c}} &=& -t \sum_{a,b,\sigma} \bar{\gamma}_{ab}
\bar{c}^{\dagger}_{a\sigma} \bar{c}^{\null}_{b\sigma}
+	\mu \sum_{a,\sigma} 
\bar{c}^{\dagger}_{a\sigma} \bar{c}^{\null}_{a\sigma} 
\nonumber \\
&& + \frac{J}{2 \mathcal{N}} \sum_{a,b,\sigma,\sigma'} \bar{\gamma}_{ab} 
\bar{c}^{\dagger}_{a\sigma} \bar{c}^{\null}_{a\sigma'} 
\bar{c}^{\dagger}_{b\sigma'} \bar{c}^{\null}_{b\sigma} .
\label{cpham} 	
\end{eqnarray}
At the large-$\mathcal{N}$ 
limit the cluster Hamiltonian (\ref{cpham}) can simply be solved exactly.
Introducing the cluster quantity 
$\bar{\chi}^{\text{c}}_{\mathbf{\bar{k}}}=(1/\mathcal{N}) \sum_{\sigma}
\bar{c}^{\dagger}_{\mathbf{\bar{k}}\sigma} \bar{c}^{\null}_{\mathbf{\bar{k}}\sigma} $
we obtain the cluster self energy
\begin{eqnarray}
	\bar{\Sigma}^{\text{c}}_{\mathbf{\bar{k}}} = - J \frac{1}{L_c}
	\sum_{\mathbf{\bar{p}}} \bar{\gamma}_{\mathbf{\bar{k}}-\mathbf{\bar{p}}} 
	\bar{\chi}^{\text{c}}_{\mathbf{\bar{p}}} , 
\end{eqnarray}
where $\bar{\gamma}_{\mathbf{\bar{k}}} = (L_c/L) \sum_{\mathbf{K}} 
\gamma(\mathbf{K}+\mathbf{\bar{k}})$.
The cluster quantity  $\bar{\chi}^{\text{c}}_{\mathbf{\bar{k}}}$ can be calculated from
the cluster Green function. We obtain
\begin{eqnarray}
\bar{\chi}^{\text{c}}_{\mathbf{\bar{k}}} &=& f(\beta \bar{E}^{\text{c}}_{\mathbf{\bar{k}}}) , 
\label{c1} \\
\bar{E}^{\text{c}}_{\mathbf{\bar{k}}} &=& -t \bar{\gamma}_{\mathbf{\bar{k}}} + \mu
-J \frac{1}{L_c} \sum_{\mathbf{\bar{p}}}
 \bar{\gamma}_{\mathbf{\bar{k}}-\mathbf{\bar{p}}}\bar{\chi}^{\text{c}}_{\mathbf{\bar{p}}} . 	
\label{c2}
\end{eqnarray}
Finally, once the cluster equations (\ref{c1})-(\ref{c2}) are solved, we obtain 
the lattice Green function of the original fermions 
\begin{eqnarray}
	G^{\text{PCPT}}(\mathbf{k},z)=\bar{G}_{\mathbf{\bar{k}}}(\mathbf{K},z) =
	\frac{1}{z+t\gamma(\mathbf{k})-\mu
	-\bar{\Sigma}^{\text{c}}_{\mathbf{\bar{k}}}} .
\end{eqnarray}
The lattice quantity $\chi^{\text{PCPT}} = (1/\mathcal{N}) \sum_{\sigma} c^{\dagger}_{i,\sigma} c^{\null}_{i+1,\sigma}$ is calculated from this lattice Green function. We obtain 
\begin{eqnarray}
\chi^{\text{PCPT}} &=& \frac{1}{2L} \sum_{\mathbf{k}} 
f(\beta E^{\text{PCPT}}(\mathbf{k})) \gamma(\mathbf{k}) , \\
E^{\text{PCPT}}(\mathbf{k}) &=& -t\gamma(\mathbf{k})+\mu+\bar{\Sigma}^{\text{c}}_{\mathbf{\bar{k}}} .	
\end{eqnarray}
So far we have obtained the quantity $\chi$ in both the CPT and PCPT, as well as
its exact value. We use it to compare the error and the convergence of the
CPT and PCPT schemes. 

\begin{figure}[t]
\begin{center}
\psfig{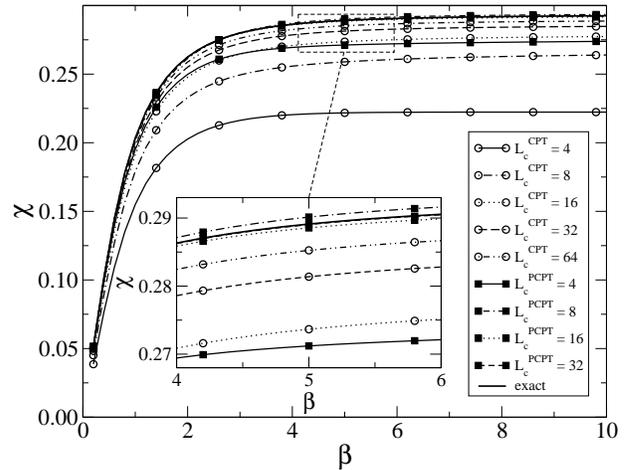}
\end{center}
\caption{\label{fig1}
The lattice quantity $\chi$ as a function of inverse temperature
$\beta$ for the CPT and PCPT with 
different cluster sizes $L_c$ ($t=J=1$, $\mu=1$).}
\end{figure}

\begin{figure}[t]
\begin{center}
\psfig{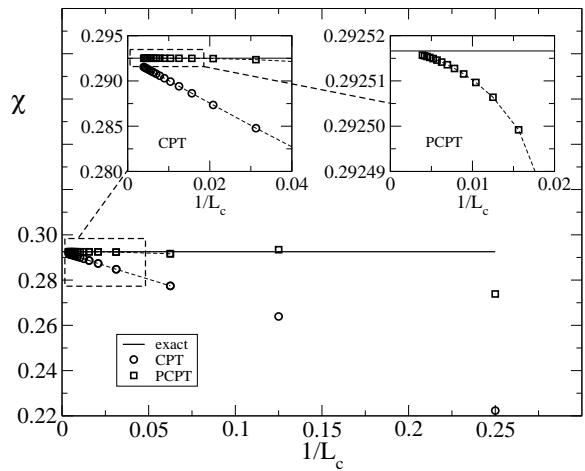}
\end{center}
\caption{\label{fig2} The lattice quantity $\chi$ as a function
of $1/L_c$ for
the CPT and PCPT at fixed temperature $\beta=10$ ($t=J=1$, $\mu=1$).}
\end{figure}

In Fig.~\ref{fig1} we plot the lattice quantity $\chi$ as a function of inverse temperature
calculated within the CPT and the PCPT. It is clear that both the CPT and PCPT
results converge well to the exact solution. The CPT results converge consistently,
whereas for small cluster sizes the PCPT results fluctuate around the exact solution.
However, the PCPT results converge faster than the CPT ones. Even at small cluster
sizes the PCPT results are closer to the exact solution than the CPT ones. In Fig.~\ref{fig2}
we plot $\chi$ as a function of $1/L_c$ at fixed temperature. It shows that the CPT
results converge linearly in $1/L_c$, whereas the PCPT results converge quadratically
in $1/L_c$. The convergence features of the CPT and PCPT are quite similar to the ones
of the CDMFT and DCA.\cite{Jarrell1,Biroli,Aryan} 
The convergence can be understood in the term of the hybridization in the locator
expansion.\cite{Jarrell1} Indeed, the convergence  of $\chi$ depends mostly on the
convergence of the coarse grained Green function (see Eq.~(\ref{cptchi}))
\begin{eqnarray}
	\hat{G}(z)= \frac{L_c}{L} \sum_{\mathbf{K}} \hat{G}(\mathbf{K},z) ,
\end{eqnarray}
where the Green function $\hat{G}(\mathbf{K},z)$ can be rewritten as
\begin{eqnarray}
	\hat{G}(\mathbf{K},z) =\big[ [\hat{G}^{\text{c}}(z)]^{-1} - \delta\hat{t}(\mathbf{K}) \big]^{-1} .
\end{eqnarray}
This is also valid for the PCPT just by replacing the coarse grained Green function
of the original fermions by its counterpart of the periodized fermions.
The coarse grained Green function can be rewritten 
in the locator expansion as\cite{Jarrell1}
\begin{eqnarray}
	\hat{G}(z)= \big[ [\hat{G}^{\text{c}}(z)]^{-1} - \hat{\Gamma}(z) \big]^{-1} ,
	\label{cg}
\end{eqnarray}
where $\hat{\Gamma}(z)$ is the hybridization function
\begin{eqnarray}
	\hat{\Gamma}(z) &=& \Big[ \hat{I} + \frac{L_c}{L} \sum_{\mathbf{K}} \delta\hat{t}(\mathbf{K}) 
	\hat{G}(\mathbf{K},z) \Big]^{-1}
	\nonumber \\
	&& \frac{L_c}{L} \sum_{\mathbf{K}} \delta\hat{t}(\mathbf{K}) \hat{G}(\mathbf{K},z)  
	\delta\hat{t}(\mathbf{K}) .
	\label{hyb}
\end{eqnarray}
The hybridization is an amplitude for fermion hopping
from a cluster into the surrounding and back again into the cluster.
It acts as the small parameter in the expansion of the coarse grained Green function
(\ref{cg}).
As in the DCA, the inter-cluster hopping
$\delta\bar{t} $ in the PCPT scales like $1/L_c$ for large cluster sizes, whereas  
in the CPT and the CDMFT it is of order $1$. The average hybridization per cluster
site scales like $1/L_c^2$ in the PCPT, and like $1/L_c$ in the CPT.\cite{Jarrell1} 
Therefore
the PCPT converges like $1/L_c^2$, whereas the CPT converges like $1/L_c$.
Note that these convergences are valid for the coarse grained Green function 
in any dimensions.  
The convergence feature
is an advantage of the PCPT in comparison with the CPT.

\section{Conclusion}
We have introduced a canonical transformation which periodizes 
fermions in the cluster space. By applying the canonical transformation
to the Hamiltonian, the periodized Hamiltonian is obtained in the presence of 
a constrain which prevents the umklapp momentum transfer from the superlattice
to the cluster space. Within the periodized Hamiltonian, the DCA can be derived
from the CDMFT. In the such way,
the DCA and the CDMFT can be unified into a single microscopic definition.
It also gives an alternative
microscopic background of the discretization of irreducible quantities on
the reciprocal space, and clarifies the approximation nature of the DCA.
We also develop the PCPT in which the wave vector within a cluster is
a conserved quantum number.
This allows to avoid the matrix inversion of the Green function
that reduces significantly the computation time when the cluster
size is large. The PCPT can work on both the direct and reciprocal spaces.
As the CPT, the PCPT is exact in  the limits 
$L_c \rightarrow \infty$, $U/t=0$, and  $t/U=0$.
It is also clarified that the small parameter of the PCPT is
$1/L_c$.  
As a benchmark the exact one dimension $1/\mathcal{N}$ model is studied. 
It turns out that the PCPT converges rapidly with corrections 
$\mathcal{O}(1/L_c^2)$, whereas the standard CPT converges with 
corrections $\mathcal{O}(1/L_c)$.

\begin{acknowledgments}

The author would like to thank Professor S.W. Kim and the Asia Pacific Center
for Theoretical Physics for the hospitality. He also acknowledges the Department
of Physics, POSTECH for sharing CPU's time.
This work was supported by
the Asia Pacific Center for Theoretical Physics.

\end{acknowledgments}

\end{document}